\documentclass{article}

\begin{document}

\title{\null\vspace*{-1truecm}  \hfill\mbox{\small CTS-IISc/6-99}\\
            \vspace*{-0.4truecm}\hfill\mbox{\tt\small quant-ph/9909082}\\
\vspace*{0.2truecm}What is Quantum Computation?}
\author{Apoorva Patel\\
{\small CTS and SERC, Indian Institute of Science, Bangalore-560012}\\
{\small E-mail: adpatel@cts.iisc.ernet.in}}
\date{{\small 9 September 1999}}
\maketitle

\begin{abstract}\noindent
Quantum computation is a rapidly progressing field today. What are its
principles? In what sense is it distinct from conventional computation?
What are its advantages and disadvantages? What type of problems can it
address? How practical is it to make a quantum computer?
I summarise some of the important concepts of quantum computation,
in an attempt to answer these questions.
A deeper understanding of them would pave the way for future development.
\end{abstract}

% Invited talk presented at the Indo-French workshop on ``Probing
% Fundamental Problems with Lasers and Cold Atoms'', IIA, Bangalore,
% India, January 1999. To appear in the proceedings.
%
% Summary of lectures presented at the ``Discussion Meeting on Quantum
% Computation'', JNCASR, Bangalore, India, July 1999.

\section{Motivation}

\rightline{\it ``Because the nature isn't classical, damn it ... ''
               ---Richard Feynman}

Let us begin by analysing what a computer is and what it actually does.
Computation is processing of information. The processing may be carried
out by a living entity or an inanimate machine, but ultimately the
processor is a physical device, and not just a mathematical construct to
implement algorithms. It follows that what is computable and what is not
is limited by the laws of physics \cite{feynmanbook}.

Traditional computer science is based on Boolean logic and algorithms.
Its basic variable is a bit, with two possible values, $0$ or $1$.
These values are represented in the computer as stable saturated states,
{\tt off} or {\tt on}. Quantum mechanics offers a new set of rules that
go beyond this classical paradigm. The basic variable is now a qubit,
represented as a normalised vector in a two dimensional complex Hilbert
space. $|0\rangle$ and $|1\rangle$ form a basis in this space, and are
physically represented as two eigenstates of a two-level quantum system.
The logic that can be implemented with such qubits is quite distinct
from Boolean logic, and this is what has made quantum computation
exciting by opening up new possibilities \cite{geometry}.

Quantum computation is thus not a question of merely implementing the
old Boolean logic rules at a different physical level with a different
set of components. We can take advantage of the novel quantum features to
devise new type of software as well as hardware. In my view, having novel
quantum devices will not make the traditional computers obsolete, rather
they would improve and enhance what is classically possible to do.

\section{Digital versus analogue computation}

Computers can be broadly classified into two types, digital and analogue,
based on the type of variables that carry information. In digital
computers the variables take only a discrete set of values, while in
analogue computers the variables belong to a continuous parameter range.

It is worthwhile to look at two specific examples, electronic computers
and living organisms (where the genetic code and the nervous system form
centres of computation). Most physical parameters performing computation
are essentially continuous, be it the voltages and currents in a circuit
or concentrations of chemicals and enzymes in a cell. A continuous
parameter can be given a sufficiently accurate digital representation by
choosing a large enough number of bits. Whether this is desirable or not
depends on the optimisation criteria of the task to be accomplished.

Major advantages of digitisation are high precision and high speed.
The former arises from the breakup of a single continuous value into a
sequence of discrete digits, while the latter is a result of a simple
instruction set. Small fluctuations and noise can be quickly corrected by
resetting the signal to the nearest known discrete value. But if a large
error does occur due to bit-flip, then its consequences typically grow
exponentially, leading the whole calculation completely astray. Also high
speed in elementary operations is accompanied by an increase in the depth
of the calculation and in power consumption.

Analogue computation has limited precision and speed. But it has a higher
tolerance against errors, in the sense that a local error would not lead
the whole computation totally off the mark. It also has the flexibility
to handle complex instructions (e.g. integration and differentiation)
reducing the depth of computation. Moreover, its low power consumption
permits high density packing of components.

Electronic computers use digital instructions on digital variables. This
strategy is optimised for high precision arithmetic and data management.
Designated components for basic Boolean operations simplify the hardware,
and make the instructions programmable into a universal machine.
Sufficient temporary memory is required to store intermediate results.

Priorities of living organisms are different, and over millions of years
they have evolved a scheme which can be dubbed digital instructions on
analogue variables. A neuron does or does not fire depending on whether
the potential from its various synapses exceeds or falls below a certain
threshold. The genetic code is composed of discrete combinations of DNA
base pairs. The instructions are often highly complex and use designated
organs for specialised tasks. They control molecular concentrations and
chemical reactions performing the desired task, and errors are weeded out
by punctuating analogue processes by digital steps. Such a combination
(even the instruction language has statistical features) is convenient
for pattern recognition problems \cite{recognition}.
For living organisms, it is far more important to guard against runaway
errors than to carry out high precision arithmetic. A brain cannot compete
in arithmetic with today's electronic computers, but in contrast the
electronic computers are no match for the brain in pattern recognition
(derivative evaluation) problems.

The reason for all this elaboration is my belief that quantum computers
are more akin to living organisms than to electronic computers. Their
best use would come from digital instructions on analogue variables,
with errors being cleaned up by digital steps punctuating analogue
processes. As we will see below, most of the research so far in quantum
computation is concentrated on how a quantum computer can beat its
classical counterpart at some specific problems in arithmetic. But the
future of quantum computation is likely to be in a different domain,
as yet unexplored.

\section{Classical information theory and computation}

Computation converts a given input to a specific output.
The process is usually deterministic, but it can be made probabilistic
and still be labeled reproducible in the sense of an ensemble.
Clearly all the information about the problem must be contained in
the input (instructions for processing data are also a type of input)
in some form or the other; the computer carries out the tedious task
of making the desired information explicit and presents it as the output.
It is not always possible to reconstruct the input from the output,
and knowledge is lost in such processes. The limiting case is the
one where all the information in the input is retained in the output,
albeit in a different form.

This description is reminiscent of the second law of thermodynamics,
with the identification of information with entropy. It is convenient
to use the language of communications to quantify information. If a
message $X$ randomly takes values $x$ with probabilities $p(x)$, then
the information conveyed by it is \cite{logbase}
\begin{equation}
  S(X) ~\equiv~ S(\{p(x)\}) ~=~ - \sum_x p(x)~\log p(x) ~~.
\end{equation}
Information is thus a measure of surprise one has upon receiving a
message. A message of repeated bits carries no information, since there
is nothing more to learn after receiving the first bit. In contrast, a
message of uniformly random bits carries maximum information, since one
has no idea of what will come next.

Shannon proved two limiting theorems based on this definition of $S(X)$.
The noiseless coding theorem gives the data compression limit:
to communicate $n$ values of $X$, one need only send only $nS(X)$ bits.
The noisy coding theorem asserts the existence of efficient error
correcting codes: over a binary symmetric channel with bit-flip error
probability $p$, a coded message of $n$ bits can transmit upto
$n (1-S(\{p,1-p\}))$ bits with an arbitrarily small error probability.

Information contained in correlations between two parts $X$ and $Y$ of
a system is described in terms of mutual entropy
\begin{equation}
  I(X:Y) = S(X) + S(Y) - S(X,Y)
         = \sum_x \sum_y p(x,y) \log [p(x,y)/p(x)p(y)] ~,
\end{equation}
where $p(x,y)$ is the joint probability for $X=x$ and $Y=y$. In absence
of correlations, $p(x,y)=p(x)p(y)$ and $I(X:Y)=0$.

The conventional paradigm for a universal digital computer is a Turing
machine. It has a finite number of internal states, a memory with unlimited
storage capacity in the form of an unbounded tape divided into cells,
a read/write head, left/right movement capability, and unconstrained
computing time at its disposal. It starts off in a certain state, looking
at the contents of a cell. Each subsequent step is determined by the current
state and the cell contents. At each step, the machine updates the current
cell contents, moves one cell to the left or right, and changes to a new
internal state. One of the states must be ``halt'' signifying the end of
computation. (Generation of random numbers is, strictly speaking, not a
computation. But with an added coin-toss instruction, a Turing machine can
perform probabilistic computation.)

Given a specific problem to solve, the programmer devises an effective
procedure or algorithm---a set of instructions for the computer to carry
out starting with some initial data. One can investigate what type of
problems can be tackled in this manner. The answer illustrates the power
of the universal Turing machine, and is summarised by the Church-Turing
hypothesis: ``Every function which would be naturally regarded as
computable can be computed by the universal Turing machine.''

From the practical point of view, it is important to find out the extent
of resources needed (hardware and time) for a specific computation, i.e.
to find out what is not just computable but also efficiently computable.
Another property of the universal Turing machine is useful here---it
can simulate any other computer with at most a polynomial overhead.
The complexity of problems can thus be classified independent of the model
of computation: those which can be solved with resources polynomial in the
input size (P), and those which require superpolynomial resources (hard).
An important subset of hard problems are those for which solutions can be
verified in polynomial time (NP). On classical computers, there exist
many superpolynomial problems: prime factorisation of large numbers,
global extremisation problems such as the travelling salesman, Boolean
circuit satisfiability problem, and so on.

G\"odel demonstrated that there exist uncomputable functions---questions
that cannot be answered at all by a consistent system of axioms and rules.
Such an incompleteness of mathematical logic/arithmetic has no place in
the physical realm---physical realisations produce physical results.
Shifting the emphasis from computation to simulation, Deutsch therefore
proposed a Church-Turing principle \cite{deutsch,feynman}:
``Every finitely realisable physical system can be simulated arbitrarily
closely by a universal quantum computer operating by finite means.''
This statement can be taken to be the definition of what a quantum
computer is and what it can do \cite{physical}.

\section{Quantum information theory}

There is no direct comparison between information content of a classical
bit that can take two discrete values and a qubit that can take any value
in a two dimensional complex Hilbert space. The best one can do is to
quantify the information of an arbitrary quantum message in units of that
of a qubit. Let us carefully consider a general quantum state and the
information that can be extracted from it.

A $n$-bit classical variable can take any of the $2^n$ discrete values.
Similarly a $n$-qubit state is a vector in the $2^n$ dimensional complex
Hilbert space. Keeping in mind that the overall phase of a quantum state
is not measurable, a normalised $n$-qubit state can be specified as
\begin{equation}
  |x\rangle ~=~ \sum_{i=0}^{2^n-1} c_i |e_i\rangle ~~,~~
  \sum_{i=0}^{2^n-1} |c_i|^2 ~=~ 1 ~~.
\end{equation}
The basis vectors of the Hilbert space can be identified with the
classical states, while the freedom to vary $c_i$ allows superposed
quantum states. With complex amplitudes $c_i$, these states are much
more general than interpolation between the extreme classical values
that can be represented by analogue devices. The superposition
principle exists in classical wave mechanics too, but it finds a
much more emphatic realisation in unusual quantum phenomena, e.g.
``Schr\"odinger's cat'' that is dead and alive at the same time.

The concept of quantum measurement adds another subtlety. It can only
be defined in probabilistic language (in the sense of an ensemble).
Consider a measurement operator with eigenstates $\{|e_i\rangle\}$.
Measuring it in the quantum state $|x\rangle$ produces an eigenvalue
$\lambda_j$ with probability $|c_j|^2$, and collapses the state to
$|e_j\rangle$ \cite{peres}. Thus it is not possible to determine all
the $c_j$ in general; the relative phase information of different
$c_j$ is lost forever. The characterisation of physical quantum
information has to be in terms of what can be extracted out of a
state, and not in terms of all the parameters that define the state.

All the properties of a quantum state are fully specified by the
density matrix $\rho$. For a pure state $|x\rangle$, it is just
the projection operator $\rho_{\rm pure}=|x\rangle\langle x|$.
The measurement process converts the pure state into a mixed state,
and $\rho$ becomes diagonal---a weighted average of pure state density
matrices
\begin{equation}
  \rho_{\rm mixed} ~=~ \sum_{j=0}^{2^n-1} |c_j|^2 |e_j\rangle\langle e_j| ~~.
\end{equation}
The accessible quantum information is then the von Neumann entropy,
\begin{equation}
S(\rho) ~=~ - {\rm Tr} (\rho \log \rho) ~~.
\end{equation}
It is zero for a pure quantum state, while it reduces to the Shannon
entropy $S(\{ |c_j|^2 \})$ for a diagonal density matrix. $S(\rho)$
also describes the quantum analogue of the noiseless coding theorem:
the quantum data compression limit for $n$ qubits chosen at random
from an ensemble of pure states is $nS(\rho)$. Dealing with ensembles
of mixed states is trickier (but inevitable when studying quantum
error correction), however, and although various bounds exist, firm
theorems regarding them are still to be established.

The most interesting feature of quantum information is contained not in
individual qubits, but in correlations amongst them, often referred to
as entanglements. Bell showed that there exist quantum entanglements
which cannot be realised by any classical probabilistic local hidden
variable theory \cite{bell}.
Such entanglements originate from complex superposition coefficients,
and exemplify physically possible tasks which no classical computer can
perform. With the reduced density matrices defined as partial traces,
the entanglement entropy for pure quantum states is
\begin{equation}
  E(X:Y) ~=~ S(\rho_X) + S(\rho_Y) - S(\rho_{XY}) ~~,~~
  \rho_A ~=~ {\rm Tr}_B (\rho_{AB}) ~~.
\end{equation}
The well-known spin singlet state is a case where the reduced density
matrices are proportional to identity and all the non-trivial properties
of the state reside in the correlation between the two spins.

\section{Quantum dynamics}

Quantum dynamics is exactly linear and unitary, unlike the non-linear
and dissipative behaviour often seen in its classical counterpart.
Time evolution of quantum states in discrete steps is conveniently
expressed using Heisenberg's matrix mechanics (in contrast to
Schr\"odinger's wave mechanics). The quantum state is represented
as a column vector with components $c_i$, and multiplying it with
a $2^n \times 2^n$ unitary matrix from left preserves its norm.
In terms of the Hermitian Hamiltonian $H$, the unitary matrix is
$U = \exp (i\int H dt)$. This dynamics is precise and reversible
($U^{-1} \equiv U^{\dag}$ reverses the evolution) \cite{nonmarkovian}.

With the superposition principle, operation of $U$ on an $n$-qubit 
state is just a single quantum step, although the corresponding matrix
multiplication amounts to $2^n$ steps on an $n$-bit classical computer
(provided that the matrix $U$ is dense enough).
This feature has been exploited to convert classically superpolynomial
problems into quantum polynomial ones. For example, before performing
an operation, set the initial state for each qubit in an $n$-qubit
string to $(|0\rangle+|1\rangle)/\sqrt{2}$. Then the final state is a
superposition of all the outcomes corresponding to every possible
classical input to the operation---there is complete parallel processing.
Of course, quantum measurement does not allow individual determination
of each outcome. But if only one property of the possible outcomes is
desired, then a cleverly designed measurement can pick it up, and we
obtain an exponential speed-up through quantum parallelism.

The linear unitary evolution makes it impossible to copy an arbitrary
unspecified state in the Hilbert space. Consider the copying operation
for two distinct states $|u\rangle$ and $|v\rangle$:
\begin{equation}
U_{copy} |u\rangle |0\rangle ~=~ |u\rangle |u\rangle ~,~
U_{copy} |v\rangle |0\rangle ~=~ |v\rangle |v\rangle ~\Longrightarrow~
\langle u|v\rangle ~=~ \langle u|v\rangle^2 ~~.
\end{equation}
This is impossible to satisfy for non-orthogonal $|u\rangle$ and
$|v\rangle$. Note that there is no problem with copying a specific state:
\begin{equation}
|x\rangle |0\rangle ~\longrightarrow\!\!\!\!\!\!\!\! /~~
|x\rangle |x\rangle ~,~ {\rm but}~~
(a|0\rangle + b|1\rangle) |0\rangle
~\mathrel{\mathop{\longrightarrow}^{\rm C-not}}~
a|0\rangle|0\rangle + b|1\rangle|1\rangle
\end{equation}
is allowed. In fact the latter is the ``controlled not'' operation,
described in more detail below, converting superposition into
entanglement. A simple extension is the allowed operation $\sum_i c_i
|e_i\rangle |0\rangle \rightarrow \sum_i c_i |e_i\rangle |f(e_i)\rangle$.

It is also impossible to extract any information from a quantum state
without disturbing it. Consider a detector in initial state
$|\psi\rangle$ interacting with a signal:
\begin{equation}
U_{int} |\psi\rangle |u\rangle ~=~ |\psi_u\rangle |u\rangle ~,~
U_{int} |\psi\rangle |v\rangle ~=~ |\psi_v\rangle |v\rangle ~\Longrightarrow~
\langle u|v\rangle ~=~ \langle\psi_u | \psi_v\rangle \langle u|v\rangle ~~.
\end{equation}
For non-orthogonal $|u\rangle$ and $|v\rangle$,
$\langle\psi_u | \psi_v\rangle = 1$ means that the final state of the
detector is the same irrespective of the state of the signal; the states
must be disturbed if something is to be learnt about them. Moreover,
a measurement operation can leave the state undisturbed, only when the
state is an eigenstate of the measurement operator (eigenstates
of Hermitian operators are orthogonal). In other situations, the act of
measurement is non-unitary, probabilistic and irreversible---the
information orthogonal to the projected eigenstate is lost.

Putting all these features together, a quantum computer can be
described as a device that is a set of qubits (a) initialised in some
known state, (b) evolved only by a succession of selected unitary
operations, and (c) measured in a specific basis. Although quantum
states and unitary evolution have entered, the classical concepts of
initialisation, deterministic control of the evolution sequence and
measurement are still part of it. These classical concepts are not
physical limitations, rather they are the limitations of our traditional
framework of knowledge acquisition through experiments.

\section{Reversibility and thermodynamics}

Historically, reversibility of computation was first investigated
to understand the limits imposed by thermodynamical laws on energy
consumption and speed of computers. According to the second law of
thermodynamics, entropy can never decrease. So the only processes
that can be fully reversed are the adiabatic ones where entropy is
held fixed. Such processes, carried out at an infinitesimal speed,
consume essentially no energy.

Most of the physical processes---in mechanics, electrodynamics,
chemical reactions, atomic physics etc.---are reversible at the
microscopic level. So a reversible model of computation only needs
implementation of elementary computational steps (i.e. logic gates)
with these processes. Landauer emphasised that increase in entropy
can be eliminated from information processing, except from the
irreversible act of erasure \cite{landauer}.
Erasure is inevitable only when readying a blank tape for input/output,
in preparation for the next computational step. A generic reversible
computer converts (input + blank tape) to (input + output). There is
no creation or destruction of information in such processing; there
is only creation of correlation/entanglement between different parts
of the system. (e.g., measurement is establishment of a perfect
correlation between the signal and the detector.) This analysis forms
the basis of a unified description of information and entropy---they
can be freely converted into each other.

Reversibility implies that nothing will happen on the average in an
equilibrium situation. To achieve something, a reversible dynamics
must start from a non-equilibrium situation or have a driving force.
Even then a reversible process will execute much like a ballistic
wave, oscillating back and forth. One must pick the correct boundary
conditions to obtain results, i.e. know precisely when and where to
start and stop.

Obviously each reversible computational element has the same number
of inputs and outputs; given the output, the input can always be
reproduced by running the computer backwards. Bennett constructed
an explicit model of reversible computation using Boolean logic
operations \cite{bennett}.
All reversible logic gates can be represented as square permutation
matrices, with each row as well as each column having only one
occurence of $1$ and the rest of the elements being $0$.
The only one bit reversible gates are ``identity'' and ``not''.
A convenient two bit reversible gate is the ``exclusive or'',
$(x,y) \rightarrow (x,x\oplus y)$. It is also called ``controlled not'',
because the second bit is flipped if the first bit is $1$ and left
unchanged otherwise. A universal set of classical reversible gates
is complete with a three bit gate, and a common choice is the 
``(controlled)$^2$ not'' gate. All these gates are their own inverses,
and any Boolean logic circuit can be constructed with these gates
as building blocks. For example, take three bits with the third one
initialised to zero. Application of C$^2$-not to the three bits,
followed by application of C-not to the first two bits produces
a simple binary adder---the second output bit is the ``sum'' of the
first two input bits while the third one is the ``carry''.

Note that the copy and measurement operations have no conflict with
reversible computation (e.g., if $y$ is initialised to $0$, C-not
just copies $x$ into $y$.). The constraints of Eqs.(8-9) are easily
solved, when the only choices for $|u\rangle$ and $|v\rangle$ are
to be the same or to be orthogonal. Models involving elastically
colliding billiard balls have been constructed to demonstrate fully
reversible copy and measurement operations. These two operations are
forbidden not by the laws of thermodynamics, but by the superposition
principle of quantum mechanics.

\section{Quantum gates and circuits}

In order to implement an algorithm, it is much more convenient to
study gate arrays that will simulate it than to find appropriate
instructions for a Turing machine. Erase, fork (requiring copy)
and feedback (requiring non-linearity) operations have no place in
quantum circuits. It has been shown that any quantum Turing machine
can be simulated by acyclic quantum gate arrays.

Quantum gates represent general unitary transformations in the Hilbert
space, describing interactions amongst qubits. Reversible Boolean
logic gates are easily generalised to quantum circuits by interpreting
them as the transformation rules for the basis states. In addition
there are gates representing continuous transformations in the Hilbert
space. Almost any two qubit quantum gate (i.e. one where the complex
phases are not rational multiples of $\pi$) is universal \cite{twobody}.
From the practical point of view, it is convenient to simplify matters
further and choose a combination of quantum and classical gates as
universal building blocks. One easy choice happens to be a general
single qubit gate (represented as a $U(2)$ matrix) and the two qubit
C-not gate \cite{barenco}.

It is important to check that the number of elementary quantum gates
needed to build up a general $n$-qubit unitary operation does not grow
exponentially with $n$. For many useful operations (including $n$-qubit
controlled gates), this number is bounded from above by a quadratic
function of $n$. Furthermore, with continuous variables and their
limited resolution in any practical implementation, we can only carry
out bounded error calculations and not arbitrary precision calculations.
This is not a limitation provided that the resource demand (e.g. the
accuracy in the specification of the complex phases) does not grow
exponentially with the reduction in error bounds. The problems that
can be tackled in this manner form the complexity class BQP---bounded
error quantum probabilistic decision problems that can be solved with
polynomial resources.

In principle, the quantum amplitudes may be encoded in space, time
or internal degrees of freedom (e.g., respectively standing waves,
travelling waves or polarisation in case of light). The quantum
elements have to couple to both the external driving force and the
desired interaction. The driving force typically carries irreversible
effects with it, and then it becomes desirable to separate it from
the quantum signal. In such a case, two different properties of the
same physical object can be used for computation, e.g. couple the
space-time degrees of freedom to the driving force and encode the
amplitudes in the internal degrees of freedom. This choice is often
also dictated by the fact that one has a better control over the
internal degrees of freedom than on the space-time ones. Thus it
has become customary to realise quantum gates in an S-matrix
framework---there are well-defined incoming and outgoing asymptotic
states, and inbetween the interaction Hamiltonian acts for a finite
amount of time. The full quantum algorithm is depicted as a
sequentially evolving network of elementary gates.

\section{Quantum algorithms}

Quantum algorithms that beat their classical counterparts exploit two
specific features: superposition and entanglement. Superposition can
transfer the complexity of the problem from a large number of sequential
steps to a large number of coherently superposed quantum states. After
parallel processing, a clever interference can extract the desired
feature from the result. Entanglement is used to create complicated
correlations that permit the desired interference. A typical quantum
algorithm starts with a highly superposed state, builds up entanglement,
and then eliminates the undesired components providing a compact result.
A uniform spread of the initial state over a large number of basis
states is easily achieved using the Walsh-Hadamard transform,
$H=(\sigma_x + \sigma_z)/\sqrt{2}$, where the $\sigma$'s are Pauli matrices.

Shor used quantum parallelism to construct a quantum Fourier transform
(QFT) \cite{shorQFT}.
The discrete Fourier transform is a unitary operation ($N=2^n$):
\begin{equation}
|x\rangle ~\longrightarrow~ {1\over\sqrt{N}} \sum_y e^{2\pi ixy/N} |y\rangle
~~,~~ x = x_{n-1}\cdot 2^{n-1} + \ldots + x_0 ~~.
\end{equation}
This is a polynomial in $\exp(2\pi ix/N)$. The classical fast Fourier
transform reduces the total number of operations from $O(N^2)$ to
$O(N \log N)$ by fully factorising this polynomial over the field of
complex numbers. In binary notation,
\begin{equation}
|x\rangle ~\longrightarrow~ {1\over\sqrt{N}}
(|0\rangle+e^{2\pi i(.x_0)}|1\rangle) \ldots
(|0\rangle+e^{2\pi i(.x_{n-1}\ldots x_0)}|1\rangle) ~~.
\end{equation}
Since each $x$-basis state goes to a factorised unentangled state,
by superposing all of them, QFT reduces the number of operations to
$O((\log N)^2)$. Only one property of the heavily superposed state
can be determined at the end; it is periodicity in Shor's algorithm
(classical randomised algorithms reduce the prime factorisation
problem to finding the period of a function). QFT is a versatile
algorithm with applications expected in many pattern finding problems.

In general, the gain extractable using quantum parallelism depends on
the structure of the problem. Grover's algorithm for finding an item
in an unsorted database is an example where the gain is quadratic
\cite{grover}.
The algorithm starts with the maximally superposed state uniformly
spread over $n$ qubits, representing $N=2^n$ items. It uses a quantum
oracle (i.e. a black box routine that gives immediate answer to a 
query) that takes the superposed state as an input, and outputs the
state after flipping the sign of the amplitude corresponding to the
desired item while leaving all the other amplitudes unaltered. The
next step is to invert all the amplitudes about their average value.
Both these steps are unitary operations, and the procedure is iterated.
After $I$ iterations, such that $(2I+1)\sin^{-1}(1/\sqrt{N})=\pi/2$,
the admixture of all the unwanted terms in the initial superposition
is eliminated and one obtains the desired item. A classical algorithm
would take $O(N/2)$ queries on average to find the desired item.
The crucial quantum ingredient here is that maximal interference
allows one to make $O(N)$ queries in $O(\sqrt{N})$ steps.
The iterations can keep cycling forever, and they have to be stopped
precisely at the right place to get the correct answer. One knows
exactly when to stop, because the overlap of the initial state with
the desired one is known, even though the desired state is unknown.

The problem of finding parity of $n$ bits is one of the toughest ones,
where the gain provided by a quantum algorithm is only a factor of two.

Quantum computers can of course be used to simulate quantum models,
i.e. systems which are idealised parts of the real world but which
still carry characteristic quantum correlations \cite{feynman}.
The important property in these problems is quantum entanglement.
Bell's theorem proves that certain types of quantum entangled states
cannot be easily simulated with classical resources. Non-local
correlations in quantum entangled states have been used to provide
quantum teleportation (a quantum state is destroyed in one place,
and after transmission of a classical signal, recreated somewhere
else). They exploit the peculiarity that, given a specific quantum
state, an operation on one of its parts can be interpreted as a
(possibly different) operation on the other part. For example,
\begin{equation}
|1\rangle (|0\rangle-|1\rangle)
~\mathrel{\mathop{\longrightarrow}^{\rm C-not}}~
|1\rangle (|1\rangle-|0\rangle) ~\equiv~
-|1\rangle(|0\rangle-|1\rangle) ~~.
\end{equation}
In the first interpretation, C-not leaves the first qubit untouched
and changes the second one, while in the second interpretation, the
first qubit undergoes a phase change and the second one remains
unaltered.

Another application of entanglement is in dense coding, where the
components of a spin-singlet state are separated, then certain
encoding steps are performed on one of the halves and it is sent to
the location of the other half. The decoding analysis of the reunited
halves provides information worth two classical bits, while only one
encoded qubit was transmitted. This example emphasises the importance
of entanglement as an information resource, and also shows that a bit
and a qubit cannot be naively equated in information content. (Quantum
teleportation is essentially the inverse operation of dense coding.)

Quantum cryptography is based on the property that extraction of even
partial quantum information from a signal leaves a signature behind.
This possibility of detecting eavesdropping is a unique quantum feature
that is absent in classical cryptography. Bennett and Brassard proposed
a protocol based on two mutually non-orthogonal sets of basis states
\cite{bb84}.
After transmission of a quantum key, an exchange of classical bits over
a public channel detects eavesdropping, eliminates noise and increases
security by distilling a smaller key composed of parity bits. The fact
that only superposition principle is used in this method has made it
easy to implement; it has already been tested over a distance of $24$
kms with existing fibre optic cables.

It should be noted that quantum algorithms are often probabilistic, but
just as in the case of classical randomised algorithms, if the success
probability is greater than half then one can hope to get to the answer
with a few trials. Also practical quantum algorithms have to be stable
against small errors, in the sense that despite round-off errors in the
continuous amplitudes, one should obtain the right answer with a bit of
extra work.

\section{Decoherence and quantum error correction}

The third law of thermodynamics forbids a finite system to have zero
entropy (or temperature). A pure quantum state having zero entropy is
therefore an idealisation. Any physical state cannot be made perfectly
pure, it will always have some interaction with its environment. Even
a reasonably well isolated quantum computer has to interact with its
surroundings for the preparation of the initial state, for receiving
instructions and for displaying the results. The inevitable external
disturbances destroy exact unitary evolution and reversibility, changing
the pure state into a mixed state. This process is called decoherence.

Physically, decoherence is a noise due to unwanted scatterings, diffusion
and localisation in the Hilbert space encoding the quantum signal.
It can be thought of as a process which entangles the quantum signal
with the environment. The evolution for the joint state of the signal
and the environment is still unitary, but the environmental degrees of
freedom are not observed. The reduced density matrix for the signal is
obtained by performing a trace over all the unobserved degrees of freedom. 
This averaging typically takes place over a large number of incoherent
variables, suppressing the off-diagonal elements of the density matrix.
The signal is reduced to a mixed state, increasing its entropy due to
neglect of information contained in the entanglement with the environment.

The precision of a practical quantum device depends on its sensitivity
to environmental disturbances, and decoherence must be controlled. Just
on dimensional grounds, the decoherence time scale due to random thermal
noise is $\hbar/kT \approx 0.76 \cdot 10^{-11}{\rm sec}/T(^\circ K)$.
It can be increased $M$-fold by combined inertia, if $M$ coherently
coupled quanta are used to represent a signal (e.g. superconductors).
It can also be increased by reducing the coupling between the quantum
state and the environment; for instance, the nuclear spins are so well
shielded by the electron cloud from their surroundings that the typical
relaxation time in NMR experiments is $O(10)$ seconds.

Generally, superposition is much more stable against decoherence than
entanglement. What is an eigenstate with a specific choice of the basis
becomes a superposed state with another choice of basis, making it
easy to manipulate superposition. Entanglement, however, is highly
fragile and must be carefully protected. Quantum algorithms that make
less use of entanglement, in space and time, are easier to implement.
(As already mentioned, quantum cryptography protocol that doesn't use
entanglement has been tested to good accuracy.) Moreover, if it is
known that some intermediate state should have a particular feature,
that can be exploited to improve the stability of an algorithm.

To make a quantum computer work longer than the decoherence time scale,
it is mandatory that checks and error corrections are built into the
system. In classical computers, bit flip errors are corrected with
redundancy, parity checks, and sophisticated Hamming codes. In the
latter case, the encoding is performed by embedding a $k$-bit code in
an $n$-bit word. By maintaining a minimum Hamming distance (i.e. the
number of bits that differ between two words) $d$ between any two
codewords, any binary vector in the $2^n$-dimensional space is within
Hamming distance $l=[(d-1)/2]$ of at most one encoded word.
Local errors flip only one bit at a time, and the code structure thus
allows upto $l$ local errors to be corrected.

These classical codes have been generalised to quantum ones, by
interpreting the words as basis vectors of the expanded Hilbert space.
With independent errors for each qubit, the quantum error operators
are direct products of $\{1, \sigma_x, \sigma_y, \sigma_z\}$ for each
qubit. Linearity of quantum mechanics then allows for their sequential
elimination. The unitary error correction step is to entangle the signal
with another string of qubits called the ancilla, such that the signal
becomes pure by transferring the error to the ancilla. The ancilla is
thereafter decoupled and discarded, restoring the signal to its proper
state. Observation of the ancilla can give a clue to the nature of the
environmental disturbances, but it gives no information regarding the
encoded state.

In a physical computer, an error may occur at several different stages---in
implementation of the logic gates, in memory, in transmission of the data,
or in the encoding/decoding steps themselves. To take care of all of them,
the computation must remain encoded throughout, from the input to the output.
In addition, a discretisation of the continuous phases helps in correcting
the errors, even at the cost of increasing the depth of computation.
Such a scheme is called fault tolerant computation \cite{shorFTCP}.
It makes arbitrary length quantum computation possible once the error
rate per qubit per gate falls below a certain finite threshold (present
estimates are about $10^{-4}-10^{-5}$).

The error correcting schemes correct only local errors, by making the
encoded information non-local. There is no way to correct global errors.
To make the errors affecting different qubits uncorrelated, we must have
the graininess (or correlation length) of the environment smaller than
the coding scale size of the qubits. This is a tough proposition to
fulfill in practice, with atomic dimensions characterising both the
environment and the qubits.

\section{Quantum hardware}

Practical requirements for a system that can be used as a quantum computer
are quite stringent: (a) The quantum degrees of freedom have to be defined
precisely (e.g. one can't have $100\pm5$ qubits doing a computation), and
there has to be a sufficient number of qubits to do a reasonable calculation;
(b) It should be possible to initialise the system in any desired state
(this may need low temperatures to remove thermal noise); (c) High degree
of isolation from environment should be available to reduce decoherence;
(d) It should be possible to subject the system to a precisely controlled
sequence of unitary transformations; (e) The measurement process should be
able to detect the state of the system with high degree of certainty. All
these features need to be implemented with a stiff tolerance criterion;
errors more than a fraction of a percent are unacceptable. Needless to say,
this is too far away in the future.

What has been realised so far is quite modest: $O(10)$ logic operations
on a few qubit systems with an accuracy of a few percent. Most quantum
components are found at the atomic scale, amongst atoms and photons.
Various proposals involve trapped ions, nuclear spins, long-lived atomic
energy levels, quantum dots, Cooper pairs, and so on. Alternatives
involving macroscopic quantum states as components (e.g. coherent states
of lasers or superconductors) have also been suggested. The important
constraint on such devices, for their quantum nature to be manifest,
is that the action for transitions between various states should be
$O(\hbar)$. At the atomic scale, movable parts are difficult to control.
So in a typical quantum device, qubits are held in place and instructions
are supplied to them by external pulses of electromagnetic radiation.

Implementation of single bit unitary operations (i.e. phase rotations)
is relatively easy, in contrast to the two bit C-not operation and
entanglement. Controlled quantum operations use a multicomponent system
whose interaction with the external fields depends on the interaction
amongst the components. (e.g. quantum energy levels can shift depending
on interactions amongst the atoms, which in turn determines whether
there will be or will not be any resonant interaction with a photon of
a particular frequency.)

At present there are two experimentally realised systems that can be
labeled quantum information processors: ions in a linear trap manipulated
using laser beams \cite{qcit},
and nuclear spin chains handled using bulk NMR methods \cite{qcnmr}.
In the first system, a string of ions are confined by electric fields in
a high vacuum. Each ion has two long-lived states which act as the two
states of a qubit. Laser beams can illuminate individual ions, allowing
transitions between the two states. The coulomb interaction between the
ions provides coupling to the vibrational modes of the ions in the trap.
The phonons in the centre-of-mass vibrational mode are excited/absorbed
by the laser photon momentum. This permits transfer of information (e.g.
the C-not operation) between any two ions. The initial state is prepared
by optical pumping and laser cooling, while the final detection is
achieved through laser fluorescence. A control over decoherence requires
submicroKelvin temperatures and high shielding from noise voltages.
That is the main limitation to enlarging the computational capacity.

The qubits in NMR experiments are the nuclear spins of individual atoms
in a molecule, coupled together by their magnetic moments. Each spin
has its characteristic resonance frequency in an oscillating magnetic
field. It can therefore be rotated by applying a pulse of appropriate
frequency and duration. Due to magnetic dipole-dipole interactions, the
resonance frequencies depend on the orientations of the neighbouring
spins, and so one can perform conditional C-not operations. NMR
experiments are carried out at room temperature with liquid compounds
containing $O(10^{23})$ molecules, however, and the desired signal has to
be cleverly extracted. The coherent quantum signal appears only as a small
deviation from an incoherent thermal background, much like a quasiparticle.
Subtracting from the density matrix the part that is proportional to
identity picks out this signal (identity remains invariant under unitary
transformations). The traceless part of the density matrix can be
initialised, manipulated and measured just like a pure quantum state,
and so it becomes the quantum processor. The main limitation here is
that with increasing number of spins the fraction of molecules
participating in a particular signal falls exponentially, making it
harder to pick out the signal from the thermal ensemble.

\section{Future directions}

The foundations of the subject of quantum computation have become well
established, but everything else required for its future growth is under
exploration. That covers quantum algorithms, logic gate operations, error
correction, understanding dynamics and control of decoherence, atomic scale
technology and worthwhile applications. I describe some possibilities below.

Reversibility of quantum computation may help in solving NP problems,
which are easy in one direction but hard in the opposite sense. Global
minimisation problems may benefit from interference effects (as seen in
Fermat's principle in wave mechanics). Simulated annealing methods may
improve due to quantum tunneling through barriers. Powerful properties
of complex numbers (analytic functions, conformal mappings) may provide
new algorithms.

Quantum field theory can extend quantum computation to allow for
creation and destruction of quanta. The natural setting for such
operations is in quantum optics. For example, the traditional double
slit experiment (or beam splitter) can be viewed as the copy operation.
It is permitted in quantum theory because the intensity of the two
copies is half the previous value. Inclusion of such particle number
non-conserving operations may speed up some algorithms \cite{self}.
% The transformations generalise from $U(N)$ to $Sp(2N)$.

Theoretical tools for handling many-body quantum entanglement are not
well developed. Its improved characterisation may produce better
implementation of quantum logic gates and possibilities to correct
correlated errors.

Though decoherence can be described as an effective process, its dynamics 
is not understood. To be able to control decoherence, one should be able
to figure out the eigenstates favoured by the environment in a given setup.

The dynamics of measurement process is not understood either, even after
several decades of quantum mechanics. Measurement is just described as a
a non-unitary projection operator in an otherwise unitary quantum theory.
Ultimately both the system and the observer are made up of quantum
building blocks, and a unified quantum description of both measurement
and decoherence must be developed. Apart from theoretical gain, it would
help in improving the detectors that operate close to the quantum limit
of observation.

For a physicist, it is of great interest to study the transition from
classical to quantum regime. Enlargement of the system from microscopic
to mesoscopic levels, and reduction of the environment from macroscopic
to mesoscopic levels, can take us there. If there is something beyond
quantum theory lurking there, it would be noticed in the struggle for
making quantum devices. We may discover new limitations of quantum
theory in trying to conquer decoherence.

Theoretical developments alone will be no good without a matching
technology. Nowadays, the race for miniaturisation of electronic
circuits is not too far away from the quantum reality of nature.
To devise new types of instruments, we must change our view-point
from scientific to technological---quantum effects are not for only
observation, we should learn how to control them for practical use.
The future is not foreseen yet, but it is definitely promising.

\section*{Acknowledgements}

In this article, I have touched upon only the basic principles of quantum
computation. More details can be found in many reviews and books on the
subject \cite{ekert,steane,rieffel,lps,preskill}.

This work was supported in part by the Rajiv Gandhi Institute of
Contemporary Studies in cooperation with the Jawaharlal Nehru Centre
for Advanced Scientific Research, Bangalore.
I am grateful to the Abdus Salam-ICTP, Trieste, for its hospitality
during the preparation of this manuscript.


\begin{thebibliography}{99}
\bibitem{geometry}{Geometry has a well-known example of what can happen
	 when one alters the basic axioms of a theoretical framework.
	 When mathematicians put aside Euclid's fifth postulate regarding
	 parallel lines, they came up with curved space geometry. This did
	 not make the old flat space geometry useless, but paved the way
	 for new applications such as general relativity.}
\bibitem{recognition}{If I meet a friend after a gap of several years, I
	 would still like to recognise him despite the fact that his face
	 may have changed somewhat. To allow for this flexibility, I don't
	 mind paying the penalty of goofing up when trying to tell apart
	 identical twins.}
\bibitem{logbase}{Shannon's choice of 2 as the base of logarithm is
 	 convenient for binary arithmetic. A different choice for the base
	 can be compensated by an appropriate value for the Boltzmann's
	 constant in physical considerations involving entropy.}
\bibitem{physical}{The label ``quantum computers'' is used here because,
	 to the best of our present knowledge, laws of quantum mechanics
	 describe the behaviour of various components of the computers
	 we design. Were still more fundamental laws of physics to be
	 discovered some day, and were we able to use them to design
	 components of a future computer, that computer will be named
	 accordingly.}
\bibitem{nonmarkovian}{Quantum dynamics can be labeled Markovian only
	 in terms of amplitudes, in the sense that the amplitude and its
	 time derivative fix the future evolution of the state. But
	 this dynamics is not Markovian in terms of probabilities, which
	 have far less information than the amplitudes. Concepts of
	 reaching an equilibrium or thermalisation cannot be defined.}
\bibitem{twobody}{It is remarkable that two-body quantum interactions
	 are sufficient for universal quantum computation. Three-body
	 interactions needed for classical reversible computation are
	 more difficult to achieve.}
\end{thebibliography}
\end{document}